\documentclass[a4paper,english,reprint]{revtex4-1}
\usepackage[T1]{fontenc}
\usepackage[latin9]{inputenc}
\usepackage{amsbsy}
\usepackage{amssymb}
\usepackage{graphicx}

\makeatletter

\pdfpageheight\paperheight
\pdfpagewidth\paperwidth

\usepackage{amsmath}

\makeatother

\usepackage{babel}
\begin{document}

\title{Free energy of formation of clusters of sulphuric acid and water
molecules determined by guided disassembly}

\author{Jamie Y. Parkinson$^{a}$, Gabriel V. Lau$^{b}$ and Ian J. Ford$^{a}$}

\address{$^{a}$Department of Physics and Astronomy, University College London,
Gower Street, London WC1E 6BT, U.K.$;{}^{b}$ Department of Chemical
Engineering, Imperial College London, South Kensington Campus, London
SW7 2AZ, U.K.}
\begin{abstract}
We evaluate the grand potential of a cluster of two molecular species,
equivalent to its free energy of formation from a binary vapour phase,
using a nonequilibrium molecular dynamics technique where guide particles,
each tethered to a molecule by a harmonic force, move apart to disassemble
a cluster into its components. The mechanical work performed in an
ensemble of trajectories is analysed using the Jarzynski equality
to obtain a free energy of disassembly, a contribution to the cluster
grand potential. We study clusters of sulphuric acid and water at
300 K, using a classical interaction scheme, and contrast two modes
of guided disassembly. In one, the cluster is broken apart through
simple pulling by the guide particles, but we find the trajectories
tend to be mechanically irreversible. In the second approach, the
guide motion and strength of tethering are modified in a way that
prises the cluster apart, a procedure that seems more reversible.
We construct a surface representing the cluster grand potential, and
identify a critical cluster for droplet nucleation under given vapour
conditions. We compare the equilibrium populations of clusters with
calculations reported by Henschel et al. {[}J. Phys. Chem. A 118,
2599 (2014){]} based on optimised quantum chemical structures.
\end{abstract}
\maketitle

\section{Introduction}

Discussions of the vapour phase often start with an ideal gas approximation,
but such a viewpoint entirely ignores the existence of molecular clusters
that form and dissolve as a consequence of the weak interactions that
exist between the constituent particles. Were it not for these ephemeral
and often disordered structures, the vapour would not easily be able
to transform itself into condensed phases when prepared at lower temperatures
or higher densities. The nucleation of such phases in these circumstances
is determined by the kinetics of the growth and decay of molecular
clusters, and considerable efforts over many years have been devoted
to understanding these processes \cite{kashchiev2000nucleation,ford2004statistical,kalikmanov2008argon,Kalikmanov-book2013}.

Direct numerical simulation of molecular clustering in a large computational
cell, necessarily with the use of empirical force fields, is increasingly
being explored (e.g. \cite{yasuoka1998molecular,Kraska06,Wedekind07,Tanaka11,Diemand13,Angelil15}),
though the expense is often very high and the application to complex
species and realistic experimental conditions is limited. A more thermodynamic
point of view is that even though nucleation is a nonequilibrium process,
the population kinetics can be framed in terms of the equilibrium
free energies of the molecular clusters that participate in the sequence
of growth and decay events, together with timescales for collisions
between clusters and monomers \cite{ford97}. Various modelling approaches
have been used to compute cluster free energies for species present
in the atmosphere, ranging from highly detailed structural studies
based on quantum chemistry \cite{kurten2008investigating,Henschel14},
to semiempirical descriptions employing the continuum properties of
the condensed phase \cite{Vehkamaki02,vehkamaki2006classical,merikanto2007new}.
Accurate modelling of molecular clusters is a particularly difficult
task, in view of their intrinsic instability and typical lack of clear
structural features such as crystalline order. Small clusters can
be liquid-like and descriptions made on the basis of their resemblance
to solids might be questionable.

Thermodynamic techniques that make no assumption of solid-like character
exist, and they have often been employed in a Monte Carlo (MC) setting
\cite{Vehkamaki00,merikanto2006}, under constraints introduced to
define an equilibrium cluster state \cite{lee1973theory}. Typically,
MC methods involve comparisons between ensembles of similar clusters,
often differing in size by one molecule, for example. A sequence of
such comparisons allows us to characterise the thermodynamic properties
of an arbitrary cluster, although constructing such a sequence can
be laborious. Molecular dynamics approaches have also been developed
\cite{Natarajan06,Tang06,Julin2008a} where fewer, or more natural
limitations are placed upon the configurational freedom available
to a cluster.

Recently, a nonequilibrium molecular dynamics (MD) method has been
developed that examines the inverse of a realistic process of cluster
formation in order to identify its thermodynamic stability \cite{tang2015free}.
The approach employs the Jarzynski equality \cite{jarzynski1996nonequilibrium},
according to which the mechanical work performed on a system during
a nonequilibrium process can be related to a change in equilibrium
Helmholtz free energy. The method, denoted cluster disassembly, can
be regarded as an MD version of thermodynamic integration \cite{Frenkel96}
and it has some intuitively appealing features \cite{tang2015free}.
External forces are used to pull a cluster apart into its constituent
molecules in a controlled or guided fashion. A variation, denoted
cluster mitosis, has also been developed where a cluster is separated
into two subclusters, again to characterise the change in free energy
associated with the process \cite{Lau15}. The methods have been successfully
tested against calculations of free energies, obtained by other techniques,
for model argon and water clusters.

In this paper we turn our attention to the disassembly of binary clusters.
We determine what is often referred to as the free energy of cluster
formation, which controls the equilibrium cluster populations in a
given mixed molecular vapour. Technically, we compute the grand potential
of a cluster for a given temperature and chemical potentials of the
species. We study clusters of sulphuric acid and water, since this
mixture has received considerable attention in connection with the
formation of aerosols in the atmosphere \cite{Kuang08,kirkby2011role,Brus11}.

In Section \ref{sec:Statistical-mechanics-of} we develop the theory
of binary cluster disassembly and describe how the free energy change
associated with such a process can be linked to equilibrium cluster
populations. In Section \ref{sec:Determining-the-free} we discuss
the MD simulations and the Jarzynski analysis that allows us to determine
the free energy of disassembly. We explore two modes of disassembly
processing and show that a `prising' technique, where molecules are
gently eased out of the cluster, has considerable advantages compared
with a more straightforward separation by steady pulling that can
give rise to mechanical `snapping' or tearing of the cluster. In Section
\ref{sec:Comparison-with-other} we compare our equilibrium cluster
populations with calculations made by Henschel \emph{et al.} \cite{Henschel14}
on the basis of minimum energy structures obtained from quantum chemistry.
In Section \ref{sec:Conclusions} we give our conclusions.

\section{Statistical mechanics of tethered binary clusters\label{sec:Statistical-mechanics-of}}

The external forces used in the method of cluster disassembly take
the form of harmonic tethers that attach each constituent molecule
of the cluster to a dedicated `guide' particle, initially placed at
the origin. In the molecular dynamics, the guides separate in a prescribed
manner and carry their tethered molecules along with them. The mechanical
work exerted by the guides on the cluster may be related to a free
energy change using the Jarzynski equality. We therefore start the
analysis of the disassembly by considering the free energy of a binary
cluster in the presence of a weak set of harmonic tethers. We relate
this to the free energy of an untethered, or free cluster, and then
to the equilibrium population of the cluster in a given vapour mixture.

\subsection{Free energies of free and tethered clusters}

We shall represent each molecule using a single position and momentum
and refer to it as a particle. The partition function of a free cluster
of $N$ particles of species 1 and $M$ particles of species 2, and
hence its Helmholtz free energy $F_{F}$, are given by
\begin{eqnarray}
Z_{F} & = & \exp[-F_{F}/k_{B}T]\nonumber \\
 & = & \frac{1}{N!M!h^{3(N+M)}}\int\prod_{j=1}^{N}\prod_{k=1}^{M}d\boldsymbol{x}_{1j}d\boldsymbol{p}_{1j}d\boldsymbol{x}_{2k}d\boldsymbol{p}_{2k}\nonumber \\
 &  & \times\exp[-H(\{\boldsymbol{x}_{1j},\boldsymbol{x}_{2k},\boldsymbol{p}_{1j},\boldsymbol{p}_{2k}\})/k_{B}T].\label{eq:Eq:1.1.1}
\end{eqnarray}
The dependence of the Hamiltonian $H$ on the momenta will not be
noted explicitly in the following, for economy of notation. We introduce
coordinates referring to the cluster centre of mass through the insertion
of unity in the form of
\begin{equation}
1\!=\!\int\!\delta\!\left(\!\boldsymbol{x}_{c}-\frac{1}{Nm_{1}+Mm_{2}}\!\left[\sum_{j=1}^{N}m_{1}\boldsymbol{x}_{1j}+\!\sum_{k=1}^{M}m_{2}\boldsymbol{x}_{2k}\right]\right)\!d\boldsymbol{x}_{c},\label{eq:1.1.3}
\end{equation}
where $m_{1}$ and $m_{2}$ are the particle masses. We hence extend
the integration through the introduction of a cluster centre of mass
variable $\boldsymbol{x}_{c}$, but insert a delta function constraint
that categorises the molecular configurations by their centre of mass
position. This is followed by a change of variables to positions of
particles with respect to $\boldsymbol{x}_{c}$, namely $\boldsymbol{x}_{1j}^{\prime}=\boldsymbol{x}_{1j}-\boldsymbol{x}_{c}$
and $\boldsymbol{x}_{2k}^{\prime}=\boldsymbol{x}_{2k}-\boldsymbol{x}_{c}$.
The partition function for the free cluster becomes
\begin{eqnarray}
 &  & Z_{F}=\frac{1}{N!M!h^{3(N+M)}}\int\prod_{j=1}^{N}\prod_{k=1}^{M}d\boldsymbol{x}_{1j}d\boldsymbol{p}_{1j}d\boldsymbol{x}_{2k}d\boldsymbol{p}_{2k}d\boldsymbol{x}_{c}\nonumber \\
 &  & \times\exp[-H(\{\boldsymbol{x}_{1j},\boldsymbol{x}_{2k}\})/k_{B}T]\nonumber \\
 &  & \times\delta\!\left(\boldsymbol{x}_{c}-\frac{1}{Nm_{1}+Mm_{2}}\!\left[\sum_{j=1}^{N}m_{1}\boldsymbol{x}_{1j}+\sum_{k=1}^{M}m_{2}\boldsymbol{x}_{2k}\right]\right)\nonumber \\
 &  & =\frac{V}{N!M!h^{3(N+M)}}\int\prod_{j=1}^{N}\prod_{k=1}^{M}d\boldsymbol{x}_{1j}^{\prime}d\boldsymbol{p}_{1j}d\boldsymbol{x}_{2k}^{\prime}d\boldsymbol{p}_{2k}\nonumber \\
 &  & \times\exp[-H(\{\boldsymbol{x}_{1j}^{\prime},\boldsymbol{x}_{2k}^{\prime}\})/k_{B}T]\nonumber \\
 &  & \times\delta\!\left(-\frac{1}{Nm_{1}+Mm_{2}}\!\left[\sum_{j=1}^{N}m_{1}\boldsymbol{x}_{1j}^{\prime}+\sum_{k=1}^{M}m_{2}\boldsymbol{x}_{2k}^{\prime}\right]\right),\,\label{eq:1.1.4}
\end{eqnarray}
which we can write as $Z_{F}=VZ_{F}^{c}$, where $V$ is the system
volume and $Z_{F}^{c}=\exp(-F_{F}^{c}/k_{B}T)$ is the partition function
for a cluster with its centre of mass fixed at the origin and $F_{F}^{c}$
is its free energy.

For a tethered cluster, the Hamiltonian will include an additional
set of harmonic potentials, each designed to hold its tethered particle
in oscillation at a frequency $\omega$, if isolated, irrespective
of mass. We have a partition function and free energy given by
\begin{eqnarray}
Z_{T} & = & \exp[-F_{T}/k_{B}T]\nonumber \\
 & = & \frac{1}{N!M!h^{3(N+M)}}\int\prod_{j=1}^{N}\prod_{k=1}^{M}d\boldsymbol{x}_{1j}d\boldsymbol{p}_{1j}d\boldsymbol{x}_{2k}d\boldsymbol{p}_{2k}\nonumber \\
 &  & \times\exp\Biggl[-\Biggl(H(\{\boldsymbol{x}_{1j},\boldsymbol{x}_{2k}\})+\sum_{j=1}^{N}\frac{1}{2}m_{1}\omega^{2}x_{1j}^{2}\nonumber \\
 &  & \qquad+\left.\left.\sum_{k=1}^{M}\frac{1}{2}m_{2}\omega^{2}x_{2k}^{2}\right)/k_{B}T\right],\label{eq:1.1.2}
\end{eqnarray}
and we then perform a transformation to centre of mass variables.
We first write\begin{widetext}
\begin{equation}
\!\!\!\sum_{j=1}^{N}\frac{1}{2}m_{1}\omega^{2}x_{1j}^{2}\!+\!\sum_{k=1}^{M}\frac{1}{2}m_{2}\omega^{2}x_{2k}^{2}\!=\!\sum_{j=1}^{N}\frac{1}{2}m_{1}\omega^{2}x_{1j}^{\prime2}\!+\!\sum_{k=1}^{M}\frac{1}{2}m_{2}\omega^{2}x_{2k}^{\prime2}\!+\frac{1}{2}(Nm_{1}+Mm_{2})\omega^{2}x_{c}^{2}\!+\omega^{2}\!\left[\sum_{j=1}^{N}m_{1}\boldsymbol{x}_{1j}^{\prime}+\!\sum_{k=1}^{M}m_{2}\boldsymbol{x}_{2k}^{\prime}\right]\cdot\boldsymbol{x}_{c},\label{eq:1.1.5}
\end{equation}
and noting that the final term can be ignored by virtue of the delta
function constraint, we obtain
\begin{eqnarray}
Z_{T} & = & \frac{1}{N!M!h^{3(N+M)}}\int\prod_{j=1}^{N}\prod_{k=1}^{M}d\boldsymbol{x}_{1j}^{\prime}d\boldsymbol{p}_{1j}d\boldsymbol{x}_{2k}^{\prime}d\boldsymbol{p}_{2k}d\boldsymbol{x}_{c}\exp\left[-\left(H\{\boldsymbol{x}_{1j}^{\prime},\boldsymbol{x}_{2k}^{\prime}\})+\Delta H\right)/k_{B}T\right]\nonumber \\
 &  & \times\delta\left(-\frac{1}{Nm_{1}+Mm_{2}}\left(\sum_{j=1}^{N}m_{1}\boldsymbol{x}_{1j}^{\prime}+\sum_{k=1}^{M}m_{2}\boldsymbol{x}_{2k}^{\prime}\right)\right)\exp\left(-\frac{1}{2}(Nm_{1}+Mm_{2})\omega^{2}x_{c}^{2}/k_{B}T\right)\nonumber \\
 & = & \frac{1}{N!M!h^{3(N+M)}}\int\prod_{j=1}^{N}\prod_{k=1}^{M}d\boldsymbol{x}_{1j}^{\prime}d\boldsymbol{p}_{1j}d\boldsymbol{x}_{2k}^{\prime}d\boldsymbol{p}_{2k}\exp\left[-H(\{\boldsymbol{x}_{1j}^{\prime},\boldsymbol{x}_{2k}^{\prime}\})/k_{B}T\right]\nonumber \\
 &  & \times\left[\frac{2\pi k_{B}T}{(Nm_{1}+Mm_{2})\omega^{2}}\right]^{3/2}\exp(-\Delta H/k_{B}T)\,\delta\left(-\frac{1}{Nm_{1}+Mm_{2}}\left(\sum_{j=1}^{N}m_{1}\boldsymbol{x}_{1j}^{\prime}+\sum_{k=1}^{M}m_{2}\boldsymbol{x}_{2k}^{\prime}\right)\right)\nonumber \\
 & = & \left[\frac{2\pi k_{B}T}{(Nm_{1}+Mm_{2})\omega^{2}}\right]^{3/2}Z_{T}^{c},\label{eq:1.1.6}
\end{eqnarray}
\end{widetext}where $\Delta H=\sum_{j=1}^{N}\frac{1}{2}m_{1}\omega^{2}x_{1j}^{\prime2}+\sum_{k=1}^{M}\frac{1}{2}m_{2}\omega^{2}x_{2k}^{\prime2}$
and $Z_{T}^{c}=\exp(-F_{T}^{c}/k_{B}T)$ is the partition function
of a cluster in the presence of tethering potentials and with its
centre of mass constrained to lie at the origin.

We shall treat the tethering terms in Eq. (\ref{eq:1.1.6}) as a perturbation.
We write  $F_{T}^{c}\approx F_{F}^{c}+\langle\Delta H\rangle_{0}$
where $F_{F}^{c}$ is the free energy of the free cluster with its
centre of mass constrained to lie at the origin, and the suffix 0
indicates that the expectation value is to be taken in an ensemble
of such clusters. We introduce single particle radial density profiles
$\rho_{NM}^{1}(x_{1}^{\prime})$ and $\rho_{NM}^{2}(x_{2}^{\prime})$
for each species in an $(N,M)$ cluster according to such an ensemble
and write
\begin{eqnarray}
\langle\Delta H\rangle_{0} & = & \frac{N}{2}\int\rho_{NM}^{1}(x_{1}^{\prime})m_{1}\omega^{2}x_{1}^{\prime2}d\boldsymbol{x}_{1}^{\prime}\nonumber \\
 &  & +\frac{M}{2}\int\rho_{NM}^{2}(x_{2}^{\prime})m_{2}\omega^{2}x_{2}^{\prime2}d\boldsymbol{x}_{2}^{\prime},\label{eq:1.1.7}
\end{eqnarray}
so that the difference in free energy between the free and tethered
cluster is:
\begin{eqnarray}
 &  & F_{F}-F_{T}=\Delta F_{T}=-k_{B}T\ln[\rho_{c}^{NM}(0)V]\label{eq:1.1.8}\\
 &  & \!\!-\frac{N}{2}\!\int\!\rho_{NM}^{1}(x_{1}^{\prime})m_{1}\omega^{2}x_{1}^{\prime2}d\boldsymbol{x}_{1}^{\prime}\!-\frac{M}{2}\!\int\!\rho_{NM}^{2}(x_{2}^{\prime})m_{2}\omega^{2}x_{2}^{\prime2}d\boldsymbol{x}_{2}^{\prime},\nonumber
\end{eqnarray}
where we have introduced $\rho_{c}^{NM}(0)=[(Nm_{1}+Mm_{2})\omega^{2}/(2\pi k_{B}T)]^{3/2}$,
which can be regarded as an inverse volume associated with the motion
of the centre of mass of the tethered cluster about the origin. The
first term on the right hand side of Eq. (\ref{eq:1.1.8}) is a correction
to the entropy of the cluster brought about by the tethering, while
the remaining terms are corrections to the energy.

\subsection{Grand potential and equilibrium cluster densities}

We now introduce the grand potential of a free $(N,M)$ cluster, defined
by
\begin{equation}
\Omega_{NM}(T,\mu_{1},\mu_{2})=F_{F}(N,M)-N\mu_{1}-M\mu_{2},\label{eq:1.1.9}
\end{equation}
where $\mu_{1}$ and $\mu_{2}$ are the chemical potentials of the
particle bath for the two species. The equilibrium densities of clusters
in a binary vapour phase can be shown \cite{ford97} to be given by
$n_{NM}=V^{-1}\exp(-\Omega_{NM}/k_{B}T)$. We could express $n_{NM}$
in terms of the density of a monomer of either species 1 or 2, so
that, for example, $n_{NM}=n_{10}\exp(-(\Omega_{NM}-\Omega_{10})/k_{B}T)$
in terms of the grand potential difference $\Omega_{NM}-\Omega_{10}=F_{F}(N,M)-F_{F}(1,0)-(N-1)\mu_{1}-M\mu_{2}$,
but it is more straightforward to proceed without introducing the
grand potential of a monomer.

Representing the particle bath as a mixture of ideal gases with readily
calculable chemical potentials, the cluster grand potential is
\begin{equation}
\Omega_{NM}=F_{F}(N,M)-Nk_{B}T\ln(\Lambda_{1}n_{1})-Mk_{B}T\ln(\Lambda_{2}n_{2}),\label{eq:1.1.10}
\end{equation}
where we introduce bath monomer densities $n_{1}=n_{10}$ and $n_{2}=n_{01}$,
and where $\Lambda_{s}=[h^{2}/(2\pi m_{s}k_{B}T)]^{3/2}$ for $s=1,2$.
Next, we consider the free energy difference $\Delta F_{D}=F_{f}-F_{T}$
between the disassembled, but still tethered, constituent particles
and the tethered cluster. $F_{f}$ is the combined free energy of
$N$ harmonic oscillators of species 1 and $M$ harmonic oscillators
of species 2, which can be written as:
\begin{equation}
F_{f}=-3k_{B}T\left[N\ln\left(\frac{k_{B}T}{\hbar\omega_{f1}}\right)+M\ln\left(\frac{k_{B}T}{\hbar\omega_{f2}}\right)\right],\label{eq:6}
\end{equation}
where $\omega_{fs}=(\kappa_{f}/m_{s})^{1/2}$ is the oscillator frequency
of species $s$, written in terms of a final value of the tethering
strength $\kappa_{f}$ and the particle mass. In the MD procedure,
the free energy of disassembly that we extract is actually $\Delta F_{{\rm MD}}=F_{f}-F_{T}^{{\rm dist}}$
since particles are distinguishable in MD. The partition functions
for distinguishable and indistinguishable particles are related through
$Z_{T}^{{\rm dist}}=N!M!Z_{T}$ and thus
\begin{equation}
\Delta F_{{\rm MD}}=\Delta F_{D}+k_{B}T\ln N!+k_{B}T\ln M!,\label{eq:8}
\end{equation}
so we can write the dimensionless grand potential of a free $(N,M)$
cluster as
\begin{eqnarray}
\Omega_{NM}/k_{B}T & = & -3\left[N\ln\left(\frac{k_{B}T}{\hbar\omega_{f1}}\right)+M\ln\left(\frac{k_{B}T}{\hbar\omega_{f2}}\right)\right]\nonumber \\
 &  & -\Delta F_{{\rm MD}}/k_{B}T+\ln(N!M!)+\Delta F_{T}/k_{B}T\nonumber \\
 &  & -N\ln(\Lambda_{1}n_{1})-M\ln(\Lambda_{2}n_{2}).\label{eq:13.1.1}
\end{eqnarray}
This gives
\begin{eqnarray}
\frac{\Omega_{NM}}{k_{B}T} & = & -N\ln(n_{1}v_{{\rm HO}})-M\ln(n_{2}v_{{\rm HO}})\nonumber \\
 &  & -\Delta F_{{\rm MD}}/k_{B}T+\ln(N!M!)-\ln\left[\rho_{c}^{NM}(0)V\right]\nonumber \\
 &  & -\frac{N}{2k_{B}T}\int\rho_{NM}^{1}(x_{1}^{\prime})\kappa_{i1}x_{1}^{\prime2}d\boldsymbol{x}_{1}^{\prime}\nonumber \\
 &  & -\frac{M}{2k_{B}T}\int\rho_{NM}^{2}(x_{2}^{\prime})\kappa_{i2}x_{2}^{\prime2}d\boldsymbol{x}_{2}^{\prime},\label{eq:1.1.12}
\end{eqnarray}
 where $v_{{\rm HO}}=(2\pi k_{B}T/\kappa_{f})^{3/2}$ with $\kappa_{f}=m_{1}\omega_{f1}^{2}=m_{2}\omega_{f2}^{2}$
is a volume representing the freedom of motion of each particle about
its guide after disassembly. The final tether strengths are taken
to be the same for both species, and the mass dependent initial tether
strengths $\kappa_{i1}$ and $\kappa_{i2}$ have been inserted into
the last two terms.

When $M=0$, the expressions reduce to those previously derived for
the single species case \cite{tang2015free}. We find that the grand
potential, correctly, has no dependence on $\hbar$ and the equilibrium
cluster density $n_{NM}=\exp(-(\Omega_{NM}+k_{B}T\ln V)/k_{B}T)$
does not depend on the system volume $V$.  Note that the equilibrium
cluster densities can be expressed in terms of the chemical potential
of a saturated vapour mixture if so desired, which is the traditional
way to proceed in nucleation theory, but here we avoid such a representation
and consider their dependence on the monomer densities $n_{1}$ and
$n_{2}$ rather than the saturated densities.

The analysis of binary clusters could easily be extended to clusters
of an arbitrary number of species, if such systems are of interest.
In the next section we turn our attention to determining the free
energy of disassembly $\Delta F_{{\rm MD}}$ using MD simulation and
the Jarzynski equality.

\section{Determining the free energy of disassembly of binary clusters \label{sec:Determining-the-free}}

\subsection{Simulation details and data analysis using the Jarzynski equality}

\emph{}

We employ a method for cluster disassembly based upon that developed
in earlier work \cite{tang2015free}. We study clusters of sulphuric
acid and water, ranging in size from dimers up to a cluster of twelve
molecules. Each molecule is harmonically tethered, initially weakly,
to one of a set of guide particles located at the origin. The tethering
force is applied to the heaviest atom in each molecule, namely the
sulphur in sulphuric acid and the oxygen in water. A preparatory MD
run of 10.5 ns duration is carried out under $NVT$ conditions in
which the system is allowed to equilibrate at 300 K in the presence
of tethering and intermolecular interactions. For the sulphuric acid
we used a set of classical interaction potentials developed by Loukonen
\emph{et al}. \cite{loukonen2010enhancing} based on a series of quantum
chemistry calculations, and the water was described by the SPC/E-F
extended simple point charge model \cite{yuet2010molecular}.

An ensemble of 1000 equilibrium configurations was selected at intervals
of 0.01 ns from the equilibrated trajectory, rejecting a very few
where a water molecule had temporarily become detached from the cluster.
A further set of simulations was then carried out, in which the equilibrated
clusters were disassembled through programmed motion of the guide
particles along with variation in the strength of the harmonic tethers.
Such a trajectory over a time interval $\tau$ provides a work of
disassembly given by
\begin{eqnarray}
W & = & \frac{1}{2}\int\limits _{0}^{\tau}\frac{d\kappa_{1}(t)}{dt}\sum_{j=1}^{N}\left[\hat{\boldsymbol{x}}_{1j}(t)-\mathbf{X}_{1j}(t)\right]^{2}dt\nonumber \\
 &  & -\int\limits _{0}^{\tau}\kappa_{1}(t)\sum_{j=1}^{N}\left[\hat{\boldsymbol{x}}_{1j}(t)-\mathbf{X}_{1j}(t)\right]\cdot\mathbf{V}_{1j}(t)\,dt\nonumber \\
 &  & +\frac{1}{2}\int\limits _{0}^{\tau}\frac{d\kappa_{2}(t)}{dt}\sum_{k=1}^{M}\left[\hat{\boldsymbol{x}}_{2k}(t)-\mathbf{X}_{2k}(t)\right]^{2}dt\nonumber \\
 &  & -\int\limits _{0}^{\tau}\kappa_{2}(t)\sum_{k=1}^{M}\left[\hat{\boldsymbol{x}}_{2k}(t)-\mathbf{X}_{2k}(t)\right]\cdot\mathbf{V}_{2k}(t)\,dt,\quad\label{eq:16}
\end{eqnarray}
where $\kappa_{s}(t)$ is the time dependent strength of the tether
attached to species $s$, $\hat{\boldsymbol{x}}_{sm}$ is the position
of the heavy atom in the $m$th molecule of species $s$, and $\mathbf{X}_{sm}$
and $\mathbf{V}_{sm}$ are the position and velocity of the associated
guide particle.

We employ the Jarzynski equality to relate the work to the shift in
free energy brought about by the change in Hamiltonian associated
with evolution from the initial to the final state of the system.
This relationship is $\Delta F_{{\rm MD}}=-k_{B}T\ln\langle\exp(-W/k_{B}T)\rangle$
where the angled brackets represent an average over the ensemble of
disassembly trajectories \cite{jarzynski1996nonequilibrium}. In principle,
the free energy change extracted in this way should not depend on
the protocol of disassembly, but in practice there can be a remnant
dependence arising from limited statistical coverage of the range
of trajectories, and we shall consider examples of such dependence
in Section \ref{sub:Optimising-the-disassembly}.

\begin{figure}
\begin{centering}
\includegraphics[width=1\columnwidth]{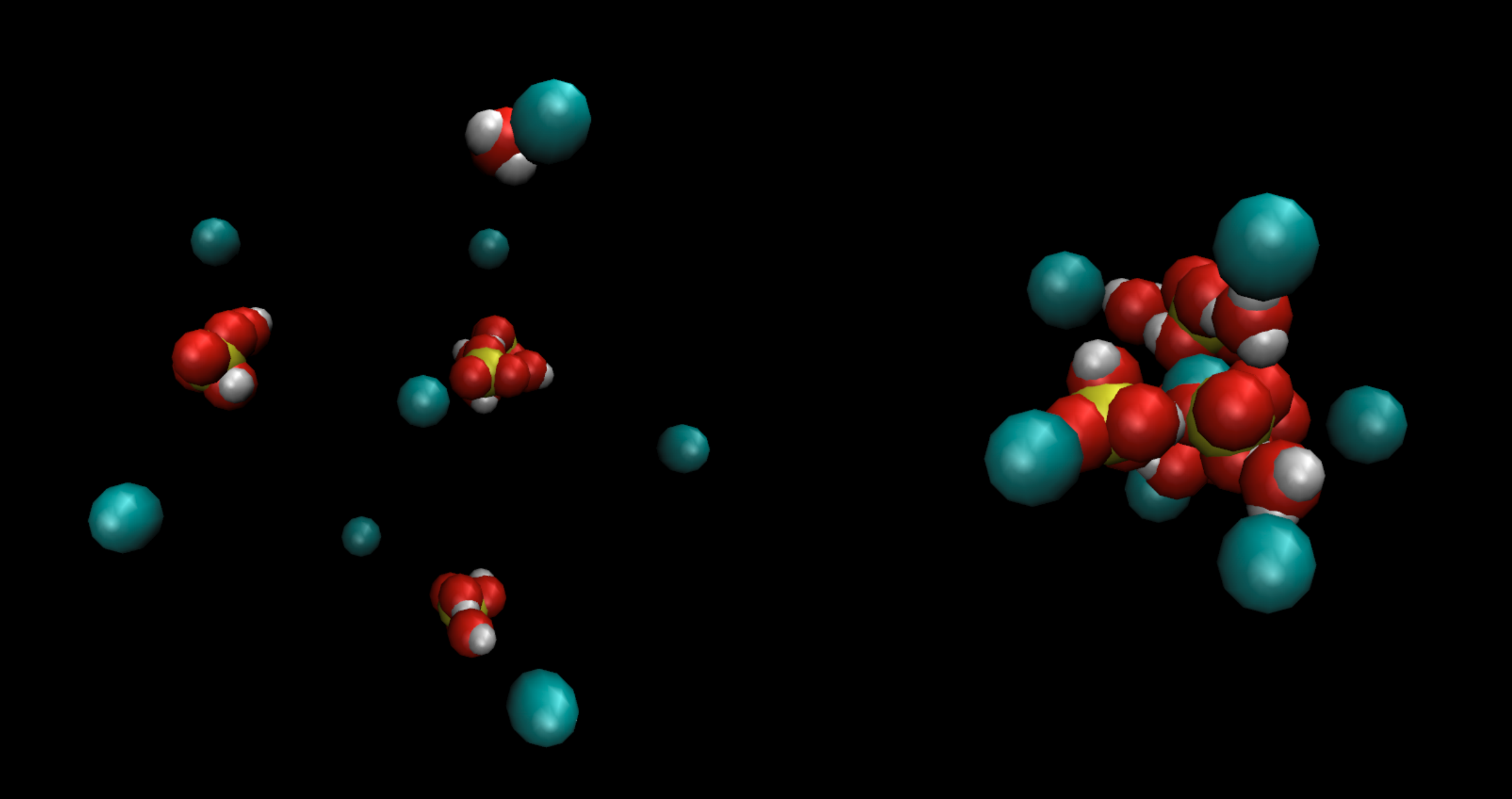}
\par\end{centering}

\caption{Illustrations of intermediate stages in the disassembly of an $N_{W}=4$,
$N_{A}=4$ cluster of sulphuric acid and water. In the image on the
left, representing the simple pulling scheme, the tethers between
the guide particles (large spheres) and the molecules are significantly
stretched, while on the right, taken from a prising simulation, molecules
are eased out of the cluster by guides situated at close range with
increasing tether strength. Movies of these disassembly trajectories
are provided in the supplemental material.\label{fig:clusterpic}}
\end{figure}

Simulations were performed using a version of the DL\_POLY molecular
dynamics package \cite{Todorov2006}, modified to implement the time
dependent harmonic tethering potentials. The molecules, but not the
guides, were coupled to a Langevin thermostat with a friction coefficient
of 0.1 ps\textsuperscript{-1}. We carried out the disassembly procedure
on a set of 46 clusters at 300 K, with the number of water molecules,
$N_{W}$, and the number of sulphuric acid molecules, $N_{A},$ ranging
from 0 to 6, excluding the cases of a monomer of either species (these
labels correspond to $N$ and $M$, respectively, in the previous
expressions). During the simulation, one of the guide particles remains
at the origin, while all others are distributed uniformly on a sphere
\cite{sloane2000tables} of time dependent radius appropriate to the
disassembly protocol. The timestep employed in all cases was 1 fs.

The initial oscillator frequencies for each species are required to
be the same, implying mass dependent initial tether strengths satisfying
$\kappa_{iW}/\kappa_{iA}=m_{W}/m_{A}$. We employ a reference initial
tether strength $\bar{\kappa}_{i}$ of 0.09 kJ mol\textsuperscript{-1}Å\textsuperscript{-2}.
This is small enough that the elevation in energy for a molecule at
the edge of the cluster is less than $k_{B}T$ \cite{tang2015free},
and slight variations have been shown elsewhere \cite{Lau15} to have
little effect on the extracted cluster free energy. We define $\kappa_{iW}=2\bar{\kappa}_{i}m_{W}/M_{WA}$
and $\kappa_{iA}=2\bar{\kappa}_{i}m_{A}/M_{WA}$, where $M_{WA}=m_{W}+m_{A}$.
Examples of cluster configurations during disassembly, together with
their guide particles, are shown in Figure \ref{fig:clusterpic}.

\subsection{Optimising the disassembly protocol\label{sub:Optimising-the-disassembly}}

We considered two modes of cluster disassembly, and implemented each
for a range of separation times to examine the convergence of the
free energy of disassembly. We first carried out a simple pulling
protocol in which the guide particles separate at a constant speed
for the duration of the simulation. The tethers start to tighten after
20\% of the total separation time and reach their maximum strength
at 80\%, in a fashion that was successfully employed in the disassembly
of argon clusters \cite{tang2015free}. The radius of the sphere on
which all but one of the guide particles are finally distributed is
25 Å, and the ultimate value of the tether strength is 0.60 kJ mol\textsuperscript{-1}Å\textsuperscript{-2}.\emph{
}The final guide particle separation was such that the tethered molecules
did not interact significantly with one another and the mean work
performed during the separation had saturated.

However it became apparent that simple pulling typically disassembled
a cluster through a sequence of abrupt `snapping' events. Molecules
showed a reluctance to separate from the cluster, indicated by an
increase in the tether length as its guide particle moved away. This
is illustrated in the left hand image in Figure \ref{fig:clusterpic}.
When the force on the molecule was strong enough to remove it from
the cluster, the guide particle had moved so far away that the extracted
molecule, after a short relaxation period characteristic of the thermostat,
had little further interaction with the cluster. Mechanically, these
extractions were typically irreversible, which also implied a thermodynamic
irreversibility, in the sense that a significant fraction of the exerted
work was converted into heat and passed to the heat bath. The contrast
with argon cluster separation in earlier work \cite{tang2015free}
was brought about by the stronger intermolecular interactions in the
present system.

Consequently, a second protocol that we call `prising' was developed
to overcome these problems. It consists of the following stages, where
$\tau$ is the separation time:
\begin{itemize}
\item $t\leq0.1\,\tau$: Motion of the guide particles to a distance of
7.5 Å from the origin, at constant initial tether strength.
\item $0.1\,\tau<t\leq0.6\,\tau$: Guide particles are held stationary,
while tethers tighten to $\kappa_{f}$ = 3.80 kJ mol\textsuperscript{-1}Å\textsuperscript{-2}.
\item $0.6\,\tau<t$: Separation of the guide particles to final positions
on a sphere of radius 10.5 Å, at constant tether strength.
\end{itemize}
The motivation for this procedure is that if the guide remains in
position as the tethers tighten, a molecule can be separated from
the cluster but remains close enough to allow interaction and also
re-attachment. The image on the right hand side of Figure \ref{fig:clusterpic}
illustrates a typical configuration from the middle stage in the prising
sequence. Once the tethers are fully tightened and the molecules prised
or eased out of the cluster, having explored a variety of ways of
doing so, the guides resume their outward motion, although we found
that little further work was performed, on average, in the third stage:
the molecules had by then been sufficiently separated (in contrast
to the 25 Å employed in the simple separation scheme with much weaker
tethering). The firm but more reversible removal of the molecules
from the cluster would be expected to lead to lower variance in the
work of disassembly\emph{. }

\begin{figure}
\begin{centering}
\includegraphics[width=1\columnwidth]{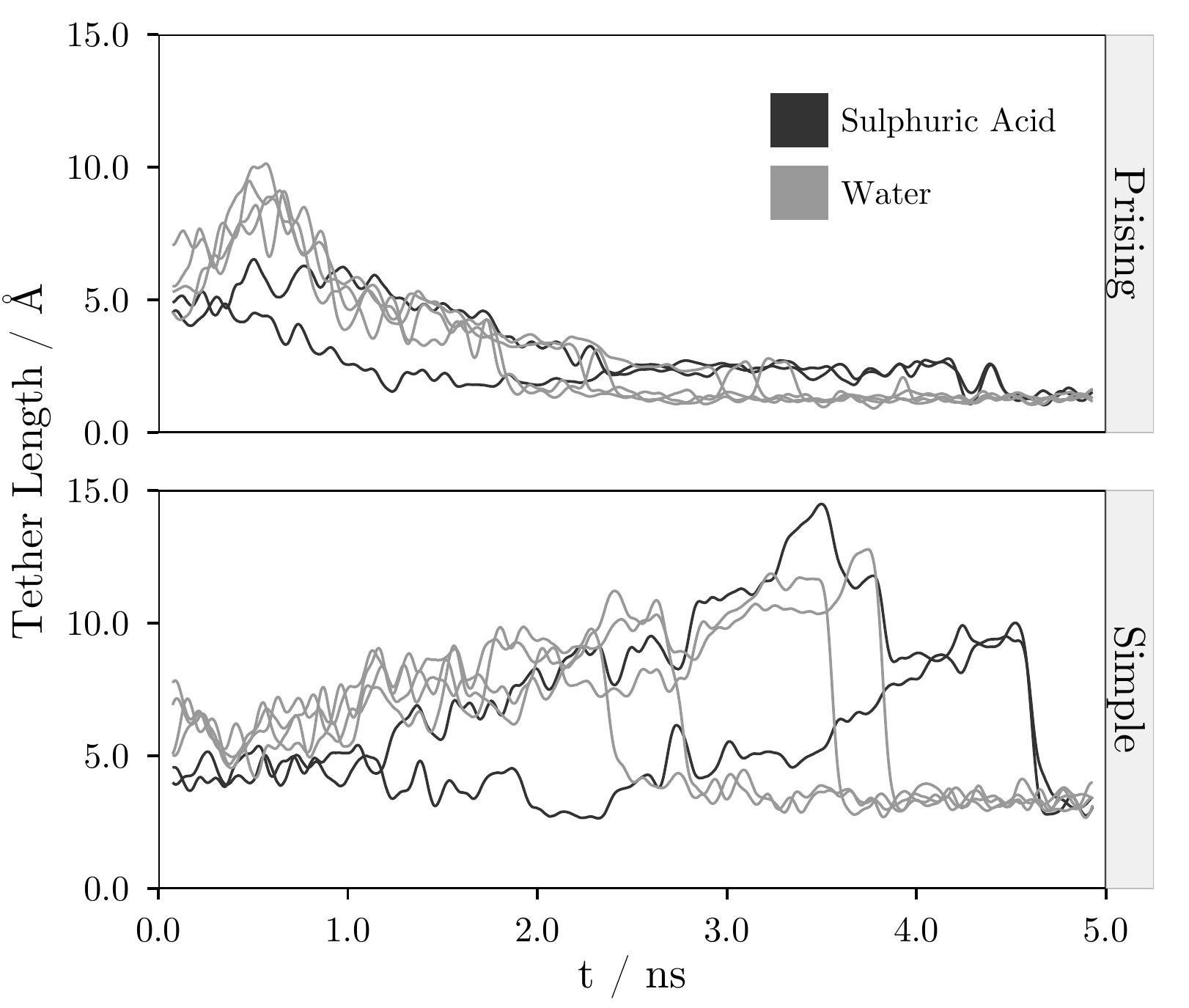}
\par\end{centering}

\caption{The evolution of tether lengths, defined as the distance between the
tethered atom and its associated guide, for the two disassembly schemes.
The cluster consists of $N_{W}=4$, $N_{A}=2$ molecules and is separated
over a time of 5 ns. Note the sharp and irreversible `snapping' of
particles out of the cluster towards their guides in the simple pulling
scheme, for example at $t\approx3.6$ ns. The data have been smoothed
using a Gaussian filter of width 0.15 ns. \label{fig:stretches}}
\end{figure}

The difference between the simple pulling and the prising protocols
is illustrated in Figure \ref{fig:stretches} for the disassembly
of a cluster of four waters and two sulphuric acid molecules over
a period of 5 ns. The lengths of a set of tethers, defined as the
distance between the guide and the atom to which it is attached, smoothed
using a Gaussian filter of width 0.15 ns to remove some of the noise,
\emph{ }are shown evolving in time for both protocols. The snapping
behaviour in the simple pulling protocol is evident in the form of
an increasing extension of the tethers followed by a rapid decrease.
For the prising case, the there is more hopping of molecules between
the guide and the cluster, and the general shortening of the tether
length with time is a consequence of the progressive tether tightening.
A movie that shows the irreversible snapping of a $N_{W}=4$, $N_{A}=4$
cluster during the simple pulling mode of disassembly is available
in the supplemental material, together with a movie of the prising
scheme.

\begin{figure}
\begin{centering}
\includegraphics[width=1\columnwidth]{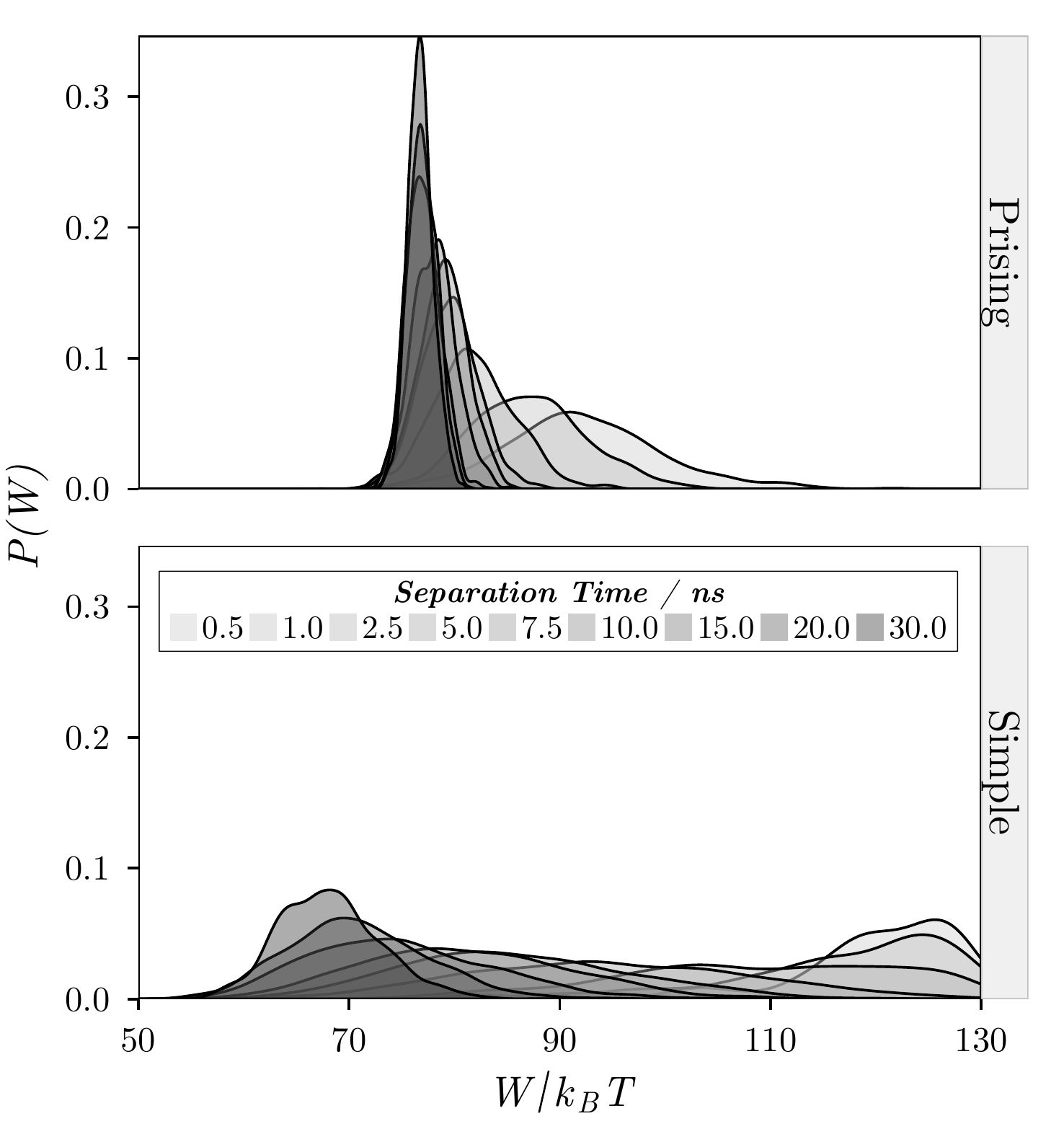}
\par\end{centering}

\caption{Kernel density estimates \cite{Scott92} for the distribution of work
$W$ (approximate probability density functions and hence denoted
$P(W)$) for the disassembly of a $N_{W}=4$, $N_{A}=2$ cluster,
for a range of separation times and for both disassembly schemes.
Note the faster convergence and the lower variance for the prising
disassembly protocol.\label{fig:densities}}
\end{figure}

\begin{figure}
\begin{centering}
\includegraphics[width=1\columnwidth]{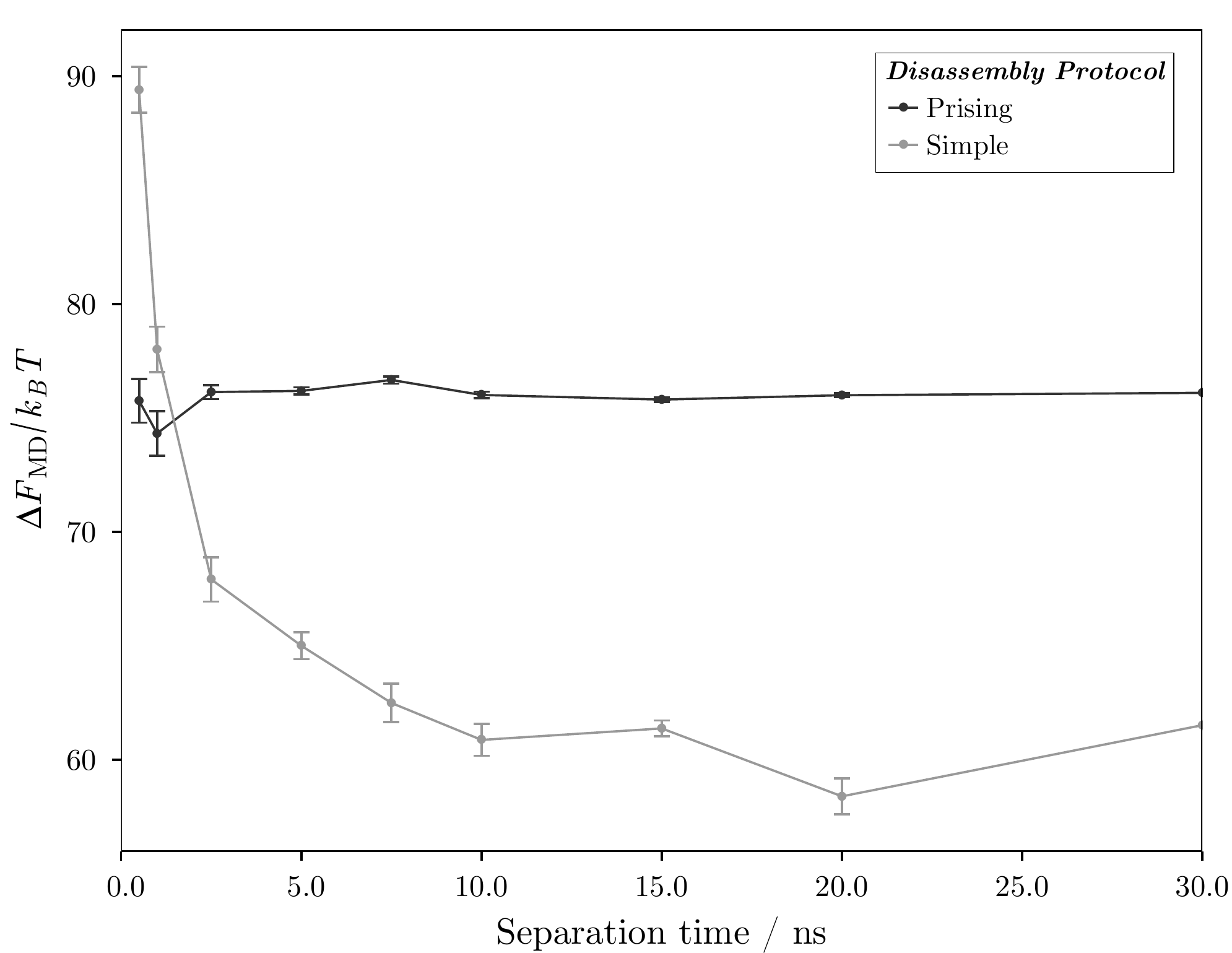}
\par\end{centering}

\caption{Estimates of the free energy of disassembly $\Delta F_{{\rm MD}}$
for a $N_{W}=4$, $N_{A}=2$ cluster, for the two disassembly schemes
and for a range of separation times. Error bars are the standard error
in the mean based on 1000 independent trajectories. The prising scheme
offers better convergence and reduced errors. Note that we should
not expect the values of $\Delta F_{{\rm MD}}$ to be the same for
each protocol, due to the difference in final tether strength. \label{fig:df}}
\end{figure}

Work distributions\emph{ }and the extracted free energies of disassembly
are shown in Figures \ref{fig:densities} and \ref{fig:df},\emph{
}respectively, for the $N_{W}=4$, $N_{A}=2$ cluster and covering
a range of separation times for both disassembly protocols. There
is a clear reduction of the variance in work and of the simulation
time required for convergence of the free energy of disassembly $\Delta F_{{\rm MD}}$
when using the prising protocol. Note that the final tether strength
in the prising protocol is higher than for simple pulling, so the
converged free energy changes are not expected to be the same for
the two approaches. The most important feature of Figure \ref{fig:df}
is that converged values of $\Delta F_{{\rm MD}}$ are obtained for
shorter simulations using the prising technique. We present grand
potentials of cluster formation in the next section based on a prising
separation time of 15 ns, although shorter times could be used without
significant loss of accuracy.

\section{Comparison with harmonic quantum chemical approach\label{sec:Comparison-with-other}}

\begin{figure}
\begin{centering}
\includegraphics[width=1\columnwidth]{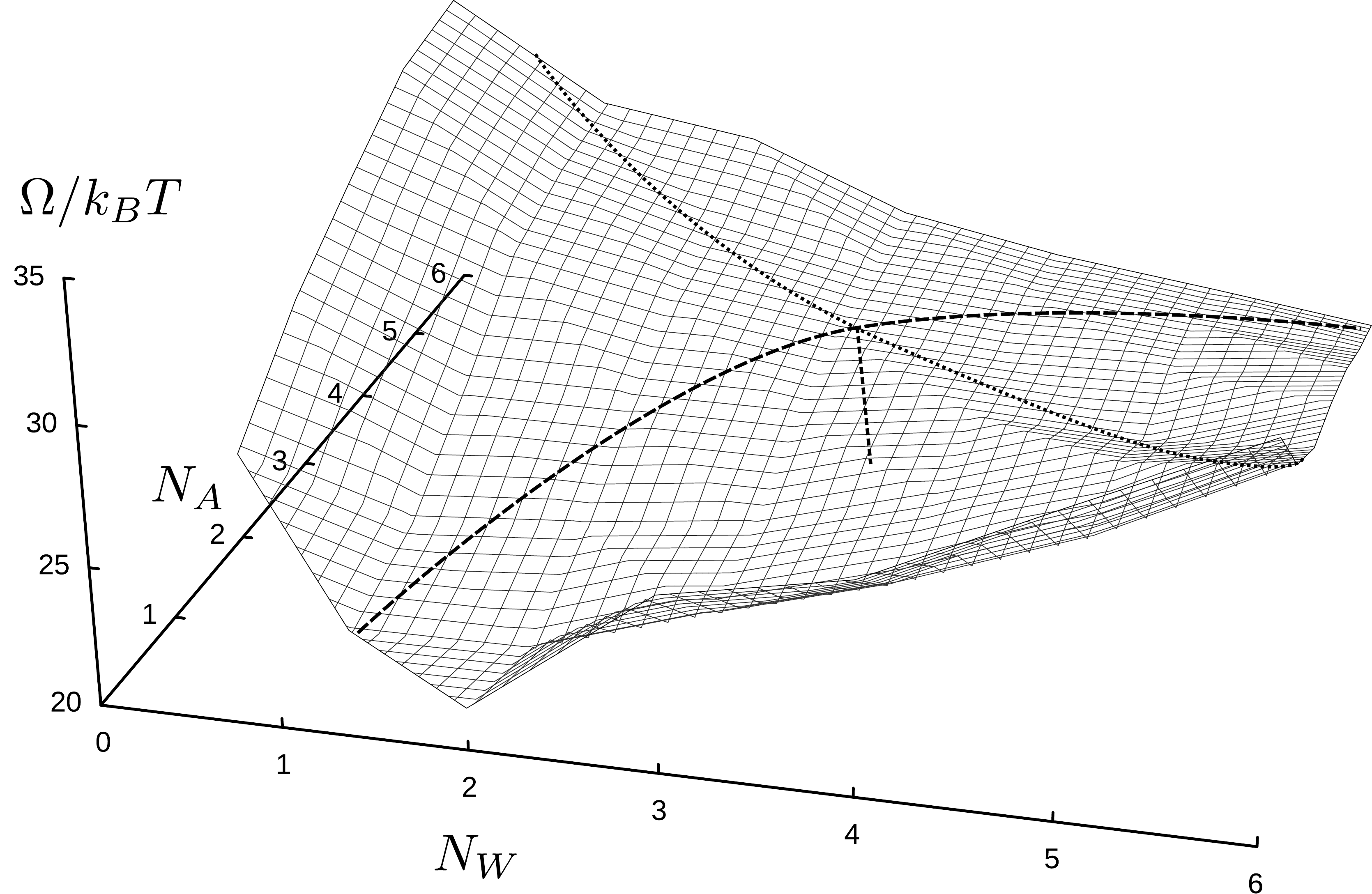}
\par\end{centering}

\caption{A surface representing the grand potential for all clusters considered,
based on sulphuric acid and water monomer densities of $2.80\times10^{-9}$
and $8.53\times10^{-8}$ Å\protect\textsuperscript{-3}, respectively,
and with $T=300$ K. The constrained equilibrium density of ($N_{W},N_{A}$)
clusters is given by $\exp(-\Omega/k_{B}T)$ in units of Å\protect\textsuperscript{-3}.
The mesh is interpolated between integer values of molecular numbers
using a 3rd-order spline interpolation. Approximate lines of steepest
descent and ascent are overlaid and the critical cluster lies at the
saddle point where they cross. An interactive version of this plot
is available as a CDF file in the supplemental material.\label{fig:surface}}

\end{figure}

We now combine the theoretical development given in Section \ref{sec:Statistical-mechanics-of}
(specifically Eq. (\ref{eq:1.1.12})) with the numerical evaluations
of the free energy of disassembly discussed in Section \ref{sec:Determining-the-free}.
We also require single particle radial density profiles $\rho_{NM}^{s}$
of a free cluster about its centre of mass, in order to evaluate the
integrals $\int\rho_{NM}^{s}(x_{s}^{\prime})x_{s}^{\prime2}d\boldsymbol{x}_{s}^{\prime}$
in Eq. (\ref{eq:1.1.12}). As an approximation, we computed these
numerically using the initial configurations of clusters in the presence
of weak tethers prior to disassembly. In Figure \ref{fig:surface}
we present the grand potential of clusters containing up to six molecules
of each species. We employ sulphuric acid and water monomer densities
of $2.804\times10^{-9}$ and $8.531\times10^{-8}$ Å\textsuperscript{-3}
respectively, ($2.804\times10^{15}$ and $8.531\times10^{16}$ cm$^{-3}$
in more commonly used units) which were chosen to make the deviations
of the surface from planarity most apparent. The surface has a saddle
point and hence a critical cluster at $N_{A}\approx4$ and $N_{W}\approx3$
for this case.

It should be emphasised that these monomer densities do not correspond
to conditions for observed particle nucleation \cite{kirkby2011role,Almeida13}.
The acid monomer density is several orders of magnitude higher than
the typical range of $10^{6}-10^{8}$ cm$^{-3}$ for sulphuric acid
in the atmosphere at 300 K \cite{Almeida13}. It is likely that atmospheric
particle nucleation proceeds with the participation of additional
molecular species, so we do not expect to find that our model reproduces
such events. Furthermore, it is quite possible that the microscopic
interactions used in this study are in need of improvement: an extended
model that permits proton transfer has recently been developed and
could be used to rectify some of its deficiencies \cite{Stinson15b}.
A similar situation was encountered in the earlier demonstration of
the cluster disassembly procedure for a single species, argon \cite{tang2015free}.
The extracted cluster thermodynamic properties were consistent with
other studies that used the same Lennard-Jones interactions, but the
implied correspondence with experimental nucleation rates for that
substance was known to be poor \cite{strey07,kalikmanov2008argon}.
Similarly, we place emphasis here on the successful implementation
of the disassembly procedure and the determination of cluster thermodynamic
properties starting from a molecular interaction scheme rather than
a presentation of realistic correlation with data. Indeed, such is
the sensitivity of cluster populations to details of the interactions,
we tend to regard a comparison with data to be informative of the
molecular interactions rather than predictive of the experimental
outcomes.

The extracted thermodynamic information can be used as the basis of
a kinetic theory of binary nucleation \cite{Reiss50,Wilemski95,vehkamaki2006classical}
but our main objective is to compare the results with a recent thermodynamic
analysis based on optimised cluster configurations and harmonic thermal
fluctuations obtained from quantum chemistry \cite{Henschel14}. We
expect differences in outcomes since the force field we use is a classical
representation of quantum mechanical interactions and might lack detail
such as a description of molecular dissociation (though this can be
remedied \cite{Stinson15b}), but equally, the harmonic approach might
not capture the correct entropic contributions to the cluster free
energy since it is fundamentally based on a solid-like conception
of each structure. We compare the approaches by computing the equilibrium
populations of clusters. Technically, these would be clusters in a
\emph{constrained} equilibrium where detailed balance is artificially
maintained between the growth and decay of clusters of all sizes and
compositions. Under the correct kinetics, cluster populations in a
stationary state of steady nucleation may be related to these equilibrium
populations.

We evaluate populations normalised by the populations of clusters
with the same number of acid molecules but no waters, namely $n_{NM}/n_{0M}$,
in order to make a direct comparison with numerical data presented
by Henschel \emph{et al}. \cite{Henschel14}. Since $n_{NM}=V^{-1}\exp(-\Omega_{NM}/k_{B}T)$
we have $n_{NM}/n_{0M}=\exp[-(\Omega_{NM}-\Omega_{0M})/k_{B}T]$,
which for $M\ne1$ involves

\begin{eqnarray}
 &  & \frac{\Omega_{NM}-\Omega_{0M}}{k_{B}T}=-N\ln(n_{1}v_{{\rm HO}})-\frac{\Delta F_{{\rm MD}}(N,M)}{k_{B}T}+\ln N!\nonumber \\
 &  & -\ln\left[\frac{\rho_{c}^{NM}(0)}{\rho_{c}^{0M}(0)}\right]-\frac{N}{2k_{B}T}\int\rho_{NM}^{1}(x_{1}^{\prime})\kappa_{i1}x_{1}^{\prime2}d\boldsymbol{x}_{1}^{\prime}\nonumber \\
 &  & -\frac{M}{2k_{B}T}\int\rho_{NM}^{2}(x_{2}^{\prime})\kappa_{i2}x_{2}^{\prime2}d\boldsymbol{x}_{2}^{\prime}+\frac{\Delta F_{{\rm MD}}(0,M)}{k_{B}T}\nonumber \\
 &  & +\frac{M}{2k_{B}T}\int\rho_{0M}^{2}(x_{2}^{\prime})\kappa_{i2}x_{2}^{\prime2}d\boldsymbol{x}_{2}^{\prime},\label{eq:21}
\end{eqnarray}
while for $M=1$ we can write $n_{N1}/n_{01}=\exp[-\Omega_{N1}/k_{B}T-\ln(n_{01}V)]$
and use

\begin{eqnarray}
 &  & \frac{\Omega_{N1}}{k_{B}T}+\ln(n_{01}V)=-N\ln(n_{01}v_{{\rm HO}})-\ln(\rho_{c}^{N1}(0)v_{{\rm HO}})\nonumber \\
 &  & -\frac{\Delta F_{{\rm MD}}(N,1)}{k_{B}T}+\ln N!-\frac{N}{2k_{B}T}\int\rho_{N1}^{1}(x_{1}^{\prime})\kappa_{i1}x_{1}^{\prime2}d\boldsymbol{x}_{1}^{\prime}\nonumber \\
 &  & -\frac{1}{2k_{B}T}\int\rho_{N1}^{2}(x_{2}^{\prime})\kappa_{i2}x_{2}^{\prime2}d\boldsymbol{x}_{2}^{\prime},\label{eq:22}
\end{eqnarray}
noting that the acid monomer density does not appear in these expressions.

\begin{figure}
\begin{centering}
\includegraphics[width=1\columnwidth]{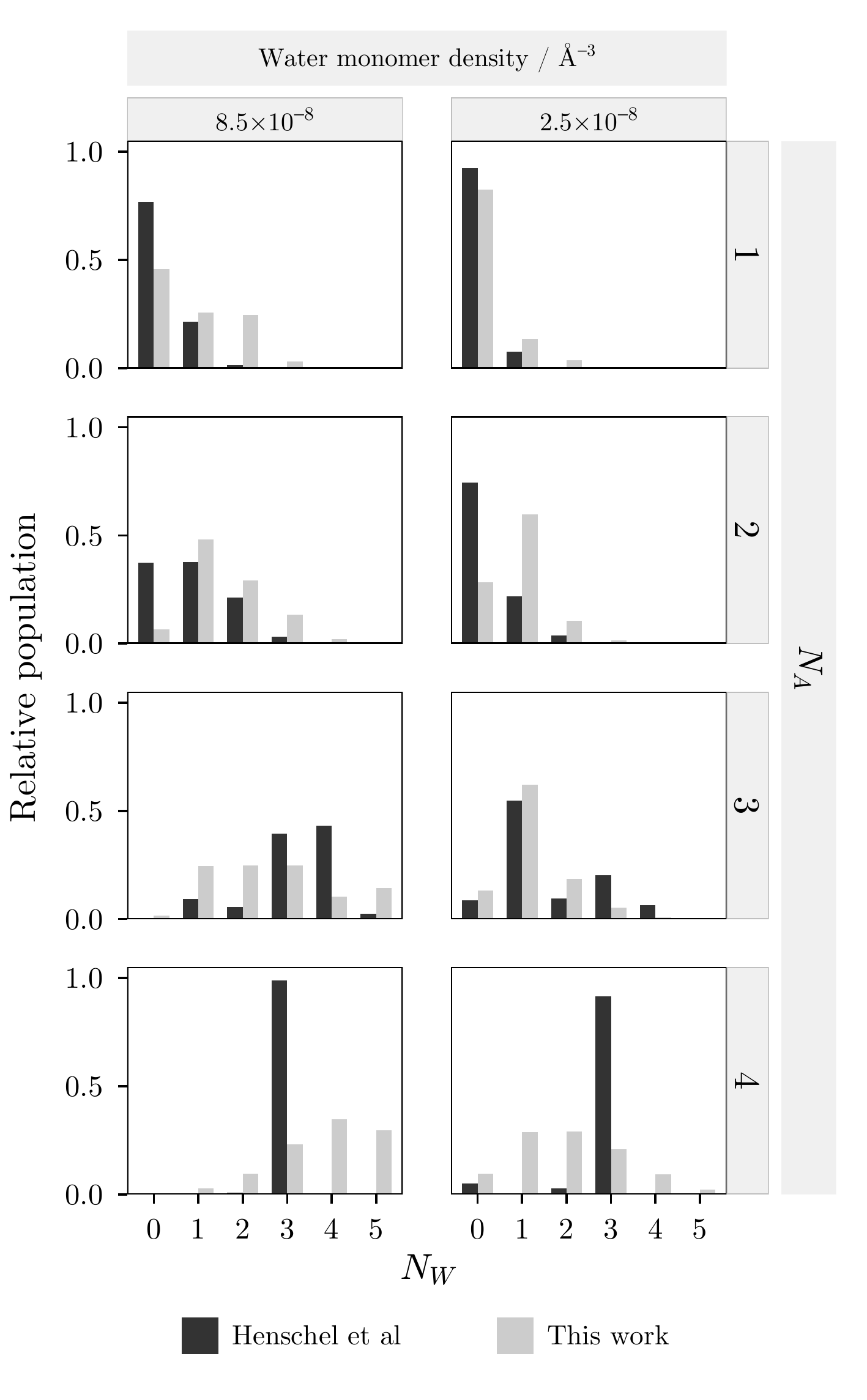}
\par\end{centering}

\caption{Populations of a variety of clusters of sulphuric acid and water at
two water monomer densities, normalised to sum to unity for each value
of $N_{A}$ (and hence independent of sulphuric acid monomer density),
calculated using our approach (light grey) and taken from Henschel
\emph{et al}. \cite{Henschel14} (dark grey). \label{fig:pops}}
\end{figure}

We use the normalised populations $n_{NM}/n_{0M}$ to construct \emph{relative}
populations $x(N_{W},N_{A})=n_{N_{W}N_{A}}/\sum_{N_{W}=0}^{5}n_{N_{W}N_{A}}$
to compare with those reported by Henschel \emph{et al}. \cite{Henschel14}
and we present these in Figure \ref{fig:pops} for two values of the
water monomer density. As expected, there are differences in detail,
but the comparison with the quantum chemical calculations is quite
reasonable, except for the acid tetramer where our approach cannot
account for the striking dominance of the trihydrated cluster in the
quantum chemical case. It remains to be seen whether the differences
can be reduced by using a better classical representation of the interactions,
or by improving the estimation of cluster entropy, or both.

\section{Conclusions\label{sec:Conclusions}}

The key to understanding first order phase transitions in complex
gaseous mixtures of precursor molecules is to compute the thermodynamic
stability of molecular clusters of various sizes and compositions.
Great strides have been made in recent years in extending quantum
chemical methods to this area, but an accurate assessment of the entropic
contributions to the relevant thermodynamic potentials, for conditions
where the clusters are liquid-like, requires an approach that goes
beyond a consideration of harmonic fluctuations, currently implying
lengthy simulation times or a reduction in the level of treatment
from quantum to classical dynamics.

Our approach is to employ a simplified force field for complex molecules
fitted to quantum chemical computations \cite{loukonen2010enhancing}
and then to use a nonequilibrium classical molecular dynamics procedure
to compute the free energy associated with the disassembly of a cluster
into its constituent molecules, avoiding a harmonic approximation
to the cluster entropy. The separation is brought about by the motion
of guide particles, each of which is tethered to one of the molecules
in the cluster. The approach offers advantages over typical Monte
Carlo methods since the assessment of the properties of a particular
cluster is more direct. We do not have to go through a lengthy comparison
of ensembles of clusters differing by only one molecule at a time;
instead we immediately refer a cluster to a system of separated, though
tethered, molecules. The approach is intrinsically a nonequilibrium
method, since it involves the mechanical manipulation of a cluster
in a finite period of time, driving the system away from equilibrium.
We exploit the Jarzynski equality \cite{jarzynski1996nonequilibrium}
to extract equilibrium free energy changes from a distribution of
nonequilibrium work, while taking care to ensure that the outcomes
are independent of the protocol of manipulation.

In this study, we have extended the method to clusters consisting
of two molecular species, noting that it could easily be generalised
to an arbitrary number. The analysis specifies the way in which the
cluster constituents should initially be tethered, and how the free
energy of disassembly should be combined with energy and entropy tethering
corrections to produce the grand potential that characterises clusters
in a binary vapour. When presented as a surface, the grand potential
conveys the intuitive idea of a thermodynamic barrier with a critical
cluster and preferred path of formation. Our approach provides a new
way to construct such a surface in a controlled and well defined fashion,
and furthermore, our presentation does not require input of the densities
of vapours in equilibrium with the condensed phase, which can complicate
the formalism.

We have demonstrated the approach by studying the important binary
system of sulphuric acid and water, and have compared our results
with those obtained recently using optimised configurations obtained
from quantum chemistry together with harmonic fluctuations. The force
field employed is simplified, and in particular does not allow proton
transfers, but more elaborate models have recently been developed
and could be employed in further studies \cite{Stinson15b}. We give
particular attention to efficiencies available through a suitable
performance of the manipulation. Instead of simply pulling the cluster
apart with soft harmonic forces, where disassembly proceeds through
a sequence of irreversible and violent snapping or tearing events,
we prise the molecules apart using guide particles that exert increasingly
strong forces while positioned at close range to the cluster. Such
a protocol favours molecular removal from the cluster in a manner
that is mechanically more reversible, which means that the free energy
of disassembly can be obtained from shorter MD simulations. We required
2-3 days of time on the multiprocessor Legion computing facility at
UCL to study 46 clusters containing different numbers of the two molecular
species.

Our computations do not capture all the dynamic subtleties of a treatment
at electronic level, since they are based on a fitted classical force
field, but they are very much faster to perform than \emph{ab initio}
methods, and should do better in assessing the entropy of a liquid-like
cluster. The comparison between our results and those of Henschel
\emph{et al}. \cite{Henschel14} is very reasonable and this study
will pave the way for further investigations based on improved force
fields \cite{Stinson15b} and a wider variety of molecular species
\cite{loukonen2010enhancing}.

\section*{Acknowledgements}

GVL was supported through a studentship in the Centre for Doctoral
Training on Theory and Simulation of Materials managed by the Department
of Physics at Imperial College, funded by the Engineering and Physical
Sciences Research Council (EPSRC) of the UK under grant number EP/G036888,
and JYP was funded by an EPSRC Undergraduate Vacation Bursary. We
thank George Jackson, Erich Müller and Patricia Hunt for their support
and we acknowledge the UCL Legion High Performance Computing Facility
(Legion@UCL), and associated support services.

%

\end{document}